\documentclass[12pt]{article}
\usepackage{amsmath}
\usepackage{amssymb,amsthm}
\usepackage{color}
\title{Instantaneous spreading versus space localization for nonrelativistic quantum systems}
\author{F.A.B.Coutinho\\
       School of Medicine, University of S\~{a}o Paulo and\\
       LIM 01 - HCFMUSP\\ 
       and\\
       W. F. Wreszinski\\
       Instituto de Fisica USP, Rua do Mat\~{a}o, s.n., Travessa R 187\\
       05508-090 S\~{a}o Paulo, Brazil}
        
\begin{document}
\maketitle
\begin{abstract}
A theorem of Hegerfeldt \cite{GCH1} establishes, for a class of quantum systems, a dichotomy between those which are permanently localized in a bounded region of space, and those exhibiting instantaneous spreading. We analyse in some detail the physical inconsistencies which follow from both of these options, and formulate which, in our view, are the basic open problems.
\end{abstract}

\section{Introduction}

Most of our experience (outside perturbation theory) with quantum mechanics concerns nonrelativistic quantum systems. This may be due to the fact that, as yet, no results on specific models of relativistic quantum field theory (rqft) in (physical) four space-time dimensions exist \cite{GlJa}. In spite of that, it is still widely believed, and there are good reasons for that \cite{Haag}, that rqft is the most fundamental physical theory. One of its most basic principles is \textbf{microcausality} (\cite{Haag}, \cite{StrWight}), which is the local (i.e., in terms of local quantum fields) formulation of Einstein causality, the limitation of the velocity of propagation of signals by $c$, the velocity of light in the vacuum:
$$
[\Phi(f),\Phi(g)] = 0
\eqno{(1)}
$$
where the fields $\Phi$ are regarded as (space-time) operator-valued distributions, when the supports of $f$ and $g$ are space-like to one another, see section 3.

Unfortunately, in spite of its enormous success, nonrelativistic quantum mechanics (nrqm) is well-known to violate Einstein causality, which is not surprising, since nrqm is supposed to derive from the the non-relativistic limit $c \to \infty$ of rqft, mathematically speaking a rather singular limit, called a group contraction, from the Poincar\'{e} to the Galilei group, first analysed by In\"{o}n\"{u} and Wigner \cite{InWi}. For some systems, such as quantum spin systems, finite group velocity follows from the Lieb-Robinson bounds \cite{LiR}, \cite{NS}: these systems are, however, approximations to nonrelativistic many-body systems. Due to the crucial importance of Einstein causality for the foundations of physics (see \cite{HMN}), it is important to understand in which precise sense nrqm is an \textbf{approximation} of a causal theory, viz., rqft.

In classical physics, acausal behavior is well-known, e.g., in connection to the diffusion equation or heat conduction problems. In both cases, these equations may be viewed as an approximation of the telegraphy equation (see \cite{Bar}, pg. 185 or \cite{MorFes}, section 7.4), and the approximations are under mathematical control. What happens in quantum theory?

For imaginary times, the heat diffusion equation becomes the Schr\"{o}dinger equation, for a free particle of mass $m$ in infinite space ($\hbar = 1$):
$$
i\frac{\partial}{\partial t} \Psi_{t} = - \frac{1}{2m} \triangle \Psi_{t} \mbox{ with } \Psi \in {\cal H}=L^{2}(\mathbf{R}^{3})
\eqno{(2.1)}
$$
The Laplacean $\triangle$ is a multiplication operator in momentum space, and the solution of (2.1) is (with $\tilde{f}$ denoting Fourier transform of $f$),
$$
\tilde{\Psi_{t}}(\vec{p}) = \exp(-it\frac{\vec{p}^{2}}{2m}) \tilde{\Psi_{0}}(\vec{p})
\eqno{(2.2)}
$$
Assuming that $\Psi_{0}$ is a $C_{0}^{\infty}(\mathbf{R}^{3})$ function, i.e., smooth with compact support, it follows from (2.2) and the ''only if'' part of the Paley-Wiener theorem (see,e.g., \cite{RSII}, Theorem IX-11) that, for any $t \ne 0$, $\Psi_{t}$ cannot be of compact support and is thus infinitely extended: one speaks of ''instantaneous spreading'' \cite{GCH}. Of course, spreading is a general phenomenon in quantum physics, and this feature is demonstrated in a varied number of situations and in several possible ways, including an exact formula for the free propagator (see, e.g., \cite{MWB}, chapter 2). The fact that the violation of Einstein causality is ''maximal'' was sharpened and made precise by Requardt \cite{Req}, who showed that, for a class of one and n body non-relativistic systems, states localized at time zero in an arbitrarily small open set of $\mathbf{R}^{n}$ are total after an arbitrarily small time.

A very nice recent review of related questions was given by Yngvason in \cite{Yng}. His theorem 1 transcribes a result proved by Perez and Wilde \cite{PeWil} (see also \cite{Yng} for additional related references), which shows that localization in terms of position operators is incompatible with causality in relativistic quantum physics. 

A different approach, which generalizes the argument after (2.2) by making a different use of analyticity, and also introduces the dichotomy mentioned in the abstract, was proposed by Hegerfeldt \cite{GCH1}:

\textbf{Theorem 1} Let $H$ be a self-adjoint operator, bounded below, on a Hilbert space ${\cal H}$. for given $\Psi_{0} \in {\cal H}$, let $\Psi_{t}, t \in \mathbf{R}$, be defined as
$$
\Psi_{t} = \exp(-iHt) \Psi_{0}
\eqno{(3)}
$$
Let $A$ be a positive operator on ${\cal H}$, $A \ge 0$, and $p_{A}$ be defined by
$$
p_{A}(t) \equiv (\Psi_{t}, A \Psi_{t})
\eqno{(4)}
$$
Then, either
$$
p_{A}(t) \ne 0 
\eqno{(5)}
$$
for almost all $t$  and the set of such $t$  is dense and open, or 
$$
p_{A}(t) \equiv 0 \mbox{ for all } t
\eqno{(6)}
$$

If, now, the probability to find a particle inside a bounded region $V$ is given by the expectation value of an operator $N(V)$, such that
$$
0 \le N(V) \le \mathbf{1}
\eqno{(7)}
$$
(e.g., $$N(V) = |\chi_{V})(\chi_{V}| \eqno{(8)}$$ , where $\chi_{V}$ is the characteristic function of $V$), and $\mathbf{1}$ the identity operator,            it follows from theorem 1, with the choice
$$
A = \mathbf{1} - N(V)
\eqno{(9)}
$$
that, if at $t=0$ a particle is strictly localized in a bounded region $V_{0}$, then, unless it remains in $V_{0}$ for all times, it cannot be strictly localized in a bounded region $V$, however large, for any finite time interval thereafter, implying a violation of Einstein causality (see also \cite{GCH1}, pg. 24, for further comments).

Our main purpose in this review is to analyse the dichotomy (5)-(6) for nonrelativistic quantum systems (in case (6) we include some relativistic systems in the final discussion in the conclusion). We start with the option given by equation (6).

\section{Systems confined to a bounded region of space in quantum theory}

Option (6) is found - with $A$  defined by (7)-(9) - in all systems restricted to lie in a finite region $V$ by \textbf{boundaries}, with a Hamiltonian $H_{V}$ in theorem 1 self-adjoint and bounded below. This includes the electromagnetic field (Casimir effect, see the conclusion), but we now concentrate on nonrelativistic quantum systems. The simplest prototype of such is the free Hamiltonian $H_{V} = -\frac{d^{2}}{dx^{2}}$, with $V=[0,L]$. The forthcoming theorem summarizes (and slightly extends) the rather detailed analysis in \cite{GarKar}, using the results in \cite{Robinson} (see the appendix of \cite{GarKar} and references given there for the standard concepts used below). Our forthcoming conclusions differ, however, from \cite{GarKar}. 

\textbf{Theorem 2.1} In the following three cases, $H_{V}$ is self-adjoint and semi-bounded:

a1) $H_{V}^{\sigma}$ on the domain 
\begin{eqnarray*}
D(H_{V}^{\sigma}) = \{\mbox{ set of absolutely continuous (a.c.) functions } \Psi \mbox{ over } [0,L]\\
\mbox{ with a.c. first derivative } \Psi^{'} \mbox{ such that } \Psi^{''} \in L^{2}(0,L)\}
\end{eqnarray*}
and satisfying the boundary condition (b.c.)
$$
\Psi^{'}(0)= \sigma_{0} \Psi(0) \mbox{ and } \Psi^{'}(L)= -\sigma_{L} \Psi(L)
\eqno{(10)}
$$
where $(\sigma_{0},\sigma_{L}) \in (\mathbf{R} \times \mathbf{R})$;

a2) $H_{V}^{\infty}$ on $D(H_{V}^{\infty})$, same as inside the brackets in a1), but with (10) replaced by
$$
\Psi(0) = \Psi(L) = 0
\eqno{(11)}
$$

a3) $H_{V}^{\theta}$, on $D(H_{V}^{\theta})$, same as inside the brackets in a1), but with (10) replaced by 
$$
\Psi(0)= \exp(i\theta) \Psi(L)
\eqno{(12)}
$$
with $\theta \in \mathbf{R}$. The case $\sigma_{0}=0$ in a1) corresponds to Neumann b.c., $\sigma_{0}>0$ to repulsive, $\sigma_{0}<0$ to attractive boundaries (see \cite{Robinson}, pg.17), with analogous statements for $\sigma_{L}$. The case a2) corresponds to setting $\sigma_{0} = -\sigma_{L} = \infty$ in (10), and is the case of impenetrable boundaries (Dirichlet b.c.). a3) is a generalization of periodic b.c.. We also have:

\textbf{Theorem 2.2} 

a) In case a1), the momentum $p=-i\frac{d}{dx}$ is not a symmetric operator;

b) In case a2), $p$ defines a closed symmetric operator $p_{\infty}$, and in case a3) it is a self-adjoint operator $p_{\theta}$, which is a self-adjoint extension of $p_{\infty}$, but for no $\theta \in \mathbf{R}$ there are functions satisfying the Dirichlet b.c. (11) in the domain $D(p_{\theta})$ of $p_{\theta}$. Furthermore, in case a3),
$$
H_{V}^{\theta} = p_{\theta}^{2} = p_{\theta}^{*} p_{\theta}
\eqno{(13)}
$$

An explicit proof of b.) may be found in \cite{GarKar}, and a.) is straightforward. We see that in cases a1) and a2) the momentum is not well-defined (as a self-adjoint operator), while it is so in case a3), in which case the expected property (13) holds.

What do we conclude fom theorems 2.1 and 2.2 (and their natural extensions to partial differential operators in higher dimensions, see \cite{Robinson}, pg. 34)? As remarked by Robinson (\cite{Robinson}, page 22), defining the probability current density $j(x)$ associated to the particle,
$$
j(x) = i(\frac{d\bar{\Psi}}{dx} \Psi(x)-\bar{\Psi}(x)\frac{d\Psi(x)}{dx})
\eqno{(14)}
$$
we see that for a1),a2), $j(0)=0=j(L)$, while, in case a3), only $j(0)=j(L)$ holds. Thus, only in cases a1),a2) the particle flux both into and out of the system is zero, corresponding to an \textbf{isolated} system, while a3) only means that all that flows in at $x=0$ flows out at $x=L$. This is the case with \textbf{periodic} b.c. (a restriction of a3)), which requires for each $\Psi$ that $\Psi(x+L)=\Psi(x)$ for all $x \in [0,L]$. we thus call a3) generalised periodic b.c.: they allow a finite system to have a momentum operator \cite{MaRo}, because, at the same time, they render the system ''infinite'' in a peculiar way, making it into a torus.  
 
We see, therefore, that the attempt to confine a quantum system in a bounded region of space by imposing on it b.c. originating from classical physics (a1),a2)) leads to physical inconsistencies, since the momentum is expected to exist as a local generator of space-translations (theorem 2.2). Generalized periodic b.c. (a3)) sometimes save the situation, for instance regarding thermodynamic quantities in statistical mechanics, which are expected (and often proven) not to depend on the boundary conditions \cite{Ru}. For expectation values and correlation functions this  need not be the case. In addition, there are situations in rqft, such as the Casimir effect, for which periodic b.c. are definitely not adequate, as we shall comment in the conclusion.

\section{The problem of instantaneous spreading}

We now come to the option given by equation (5). Using theorem 1, Hegerfeldt (see \cite{GCH} and references given there) proposed to analyse a two-atom model suggested by Fermi \cite{Fermi} to check finite propagation speed in quantum electrodynamics, with $H$ the Hamiltonian, $A=A_{e_{B}}$, the probability that atom $B$, initially in the ground state, is excited by a photon resulting from the decay of atom $A$, initially in an excited state, and $\Psi_{0}$ denoting the initial physical state of the system $A-B$. The conclusion is that $B$ is either immediately excited with nonzero probability or never. This conclusion was challenged by Buchholz and Yngvason \cite{BYng} in a beautiful and subtle analysis, in which they concluded that there are no causality problems for the Fermi system in a full description of the system by rqft. One important point raised in \cite{BYng} is that (4),(5) with $A$ positive is not an adequate criterion to investigate causality in rqft, as shown by the simple counterexample of the state $(\Psi_{0}, \cdot \Psi_{0})$ equal to the vacuum state, for which $p_{A}(t)$ is \textbf{always} nonzero for $A$ positive space-localized, by the Reeh-Schlieder theorem (see, e.g., \cite{Ar}, pg. 101). It is perhaps worth noting that a non-perturbative rqft description of the two-atom system is not known, but the authors \cite{BYng} relied on the general principles of rqft (\cite{Haag}, \cite{Ar}).

It follows from the above that instantaneous spreading would cease to be an obstacle to the physical consistency of nonrelativistic quantum mechanics if it could be shown that the latter is an approximation, in a suitable precise sense, to rqft. In this review we expand on a discussion by C. J\"{a}kel and one of us \cite{JaWre} on this matter. In a not yet precise fashion (but see Lemma 2), one might propose as approximation criterion

\textbf{Proposal C} Nonrelativistic ground state expectation values are ''close'' to the corresponding relativistic vacuum expectation values when certain physical parameters are ''small''.

For an atom, e.g. hydrogen, in interaction with the electromagnetic field in its ground state, one such parameter is the ratio between the mean velocity of  the electron in the ground state and $c$, which is of order of the fine structure constant. It is clear that the Dirac atom, with a potential, is not a fully relativistic system, and therefore not a candidate to solve the Einstein causality problems in the manner proposed in \cite{BYng}: thus, the well-known relativistic corrections \cite{JJS} do not solve the causality issue as sketched above. Perturbative quantum electrodynamics, in spite of its great success, does not offer a solution either: for instance, the relativistic Lamb shift relies strongly on Bethe's nonrelativistic treatment (see, e.g., \cite{JJS}, pg. 292). 

We now attempt to make proposal C precise, and, at the same time, show some results relating relativistic and nonrelativistic systems which are not found in this form in the textbook literature. We take as the nonrelativistic systems, formulated in Fock space, the symmetric Fock space for Bosons, ${\cal F}_{s}({\cal H})$ which we simply denote by ${\cal F}$ (see \cite{MaRo} for a nice textbook presentation) and there the state $\omega_{\Psi_{0}}=(\Psi_{0}, \cdot \Psi_{0})$. The observables will be functionals of the nonrelativistic free quantum fields at time zero:
$$
\Phi(\vec{x}) = \phi(\vec{x}) + \phi^{*}(\vec{x})
\eqno{(15.1)}
$$
where $*$ denotes hermitian conjugate and
$$
\phi(\vec{x}) = \frac{1}{(2\pi)^{3/2}(2m_{0})^{1/2}}\int d\vec{k} a(\vec{k})\exp(-i\vec{k}\cdot \vec{x})
\eqno{(15.2)}
$$
and the canonically conjugate momenta
$$
\Pi(\vec{x}) = \pi(\vec{x}) + \pi^{*}(\vec{x})
\eqno{(16.1)}
$$
where
$$
\pi(\vec{x}) = -\frac{i(2m_{0})^{1/2}}{(2\pi)^{3/2}}\int d\vec{k} a(\vec{k})\exp(i\vec{k}\cdot \vec{x}) 
\eqno{(16.2)}
$$
Above, $a, a^{*}$ are annihilation-creation operators satisfying
$$
[a(\vec{k}),a^{*}(\vec{l})] = \delta(\vec{k}-\vec{l})
\eqno{(17)}
$$
It is more adequate, both mathematically and physically (\cite{MaRo},\cite{RSII}) to use the smeared fields
$$
\Phi(f) = \int d\vec{x} f(\vec{x}) \Phi(\vec{x})
\eqno{(18)}
$$
and
$$
\Pi(g) = \int d\vec{x} g(\vec{x}) Pi(\vec{x})
\eqno{(19)}
$$
i.e., to consider $\Phi,\Pi$ as operator-valued distributions, satisfyind the canonical commutation relations (CCR)
$$
[\Phi(f),\Pi(g)] = i(f,g)
\eqno{(20)}
$$
on a suitable dense set (\cite{RSII}, pg. 232), with
$$
(f,g) = \int d\vec{x} \bar{f}(\vec{x}) g(\vec{x})
\eqno{(21)}
$$
for $f,g \in {\cal S}(\mathbf{R}^{3})$, the Schwarz space \cite{RSII}. For the free relativistic quantum system, the corresponding state is again the no-particle state $\omega_{\Psi_{0}}$, the observables (functionals of) the relativistic free quantum fields
$$
\Phi_{r}(\vec{x}) = \phi_{r}(\vec{x}) + \phi_{r}^{*}(\vec{x})
\eqno{(22.1)}
$$
where
$$ 
\phi_{r}(\vec{x}) = \frac{c}{(2\pi)^{3/2}}\int d\vec{k}\frac{1}{(2\omega_{\vec{k}}^{c})^{1/2}} a(\vec{k})\exp(-i\vec{k}\cdot \vec{x})
\eqno{(22.2)}
$$
and the canonically conjugate momentum
$$
\Pi_{r}(\vec{x}) = \pi_{r}(\vec{x}) + \pi_{r}^{*}(\vec{x})
\eqno{(23.1)}
$$
with
$$
\pi_{r}(\vec{x}) = -\frac{i}{(2\pi)^{3/2}c}\int d\vec{k}(2\omega_{\vec{k}}^{c})^{1/2} a(\vec{k})\exp(i\vec{k}\cdot \vec{x})
\eqno{(23.2)}
$$
It is convenient to consider the CCR in the Weyl form
$$
\exp(i\Pi(f))\exp(i\Phi(g))=\exp(i\Phi(g))\exp(i\Pi(f))\exp(-i(f,g))
\eqno{(24)}
$$
for $f,g \in {\cal S}_{\mathbf{R}}(\mathbf{R}^{3})$, the Schwarz space of real-valued functions on $\mathbf{R}^{3}$. Above,
$$
\omega_{\vec{k}}^{c} \equiv (c^{2}\vec{k}^{2}+m_{0}^{2}c^{4})^{1/2}
\eqno{(25)}
$$
with $m_{0}$ the ''bare mass'' of the particles. We write
$$
a(f) = (2m_{0})^{1/2}[\Phi(f)+i \Pi(f)]
\eqno{(26)}
$$
and similarly for $a^{*}(f),a_{r}(f),a_{r}^{*}(f)$. The zero-particle vector $\Psi_{0} \in {\cal F}$ is such that
$$
a(f)\Psi_{0} = 0 \mbox{ for all } f \in {\cal S}(\mathbf{R}^{3})
\eqno{(27)}
$$
and similarly for $a_{r}(f)$. We assume that there exists a continuous unitary representation $U(\vec{a},R)$ of the Euclidean group $\vec{x} \to R\vec{x}+\vec{a}$ on ${\cal F}$ with $R$ a rotation and $\vec{a}$ a translation, s.t.
$$
U(\vec{a},R) a(f) U(\vec{a},R)^{-1} = a(f_{\vec{a},R})
\eqno{(28)}
$$
with
$$
f_{\vec{a},R}(\vec{x})= f(R^{-1}(\vec{x}-\vec{a})  
\eqno{(29)}
$$
The following lemma is fundamental:

\textbf{Lemma 1} The no-particle state $\Psi_{0}$ is the unique state invariant under $U(\vec{a},R)$.

\textbf{Proof} By (27),(28),
$$
0 = U(\vec{a},R) a(f) \Psi_{0} = a(f_{\vec{a},R})U(\vec{a},R) \Psi_{0}
\eqno{(30)}
$$
Since, for all $f \in {\cal S}(\mathbf{R}^{3})$, $U(\vec{a},R) \Psi_{0}$ is also a no-particle state by (30), it follows that
$$
U(\vec{a},R) \Psi_{0} = \lambda(\vec{a},R) \Psi_{0}
\eqno{(31)}
$$
with $|\lambda|=1$, and the $\lambda$ form a one-dimensional representation of the Euclidean group. Since the Euclidean group posesses only the trivial one-dimensional representation, we conclude that
$$
U(\vec{a},R) \Psi_{0} = \Psi_{0}
$$
i.e., $\Psi_{0}$ is necessarily a Euclidean invariant state (As Wightman observes \cite{Wight}, this is \textbf{not} assumed when one writes (28)!). In the case of the free (relativistic or nonrelativistic) field, the cluster property of the two-point function (a corollary of the Riemann-Lebesgue lemma, see, e.g., \cite{MWB}, Lemma 3.8) implies, together with von Neumann's ergodic theorem (see, again, e.g., \cite{MWB}, Theorem A.2), that $\Psi_{0}$ is the unique state invariant under all space translations and, thus, the unique state invariant under all $U(\vec{a},R)$. q.e.d.

Lemma 1 is the main ingredient of

\textbf{Theorem 3} The representations of the Weyl CCR (24) $(\Phi,\Pi)$ and $(\Phi_{r},\Pi_{r})$ are unitarily inequivalent.

\textbf{Proof} The proof of this theorem follows from (\cite{RSII}, Theorem X.46), the inequivalence of the Weyl CCR for different masses $m_{1}$ and $m_{2}$, by identifying $\Phi_{m_{1}}$ with $\Phi$ and $\Phi_{m_{2}}$ with $\Phi_{r}$ (and similarly for the $\Pi$). Let $G(R,\vec{a})$ (resp. $G_{r}(R,\vec{a})$) be the representatives of the Euclidean group leaving $(\Phi,\Pi)$ (resp.$(\Phi_{r},\Pi_{r})$) invariant. We assume that there exists a unitary map $T$ on ${\cal F}$ which satisfies $$T \exp(i\Phi(f))T^{-1} = \exp(i\Phi_{r}(f)) \eqno{(32)}$$ and $$T \exp(i\Pi(f))T^{-1} = \exp(i\Pi_{r}(f) \eqno{(33)}$$. Exactly as in \cite{RSII}, Theorem X.46, pg.234, this leads to 
$$
TG(R,\vec{a})T^{-1} = G_{r}(R,\vec{a})
\eqno{(34)}
$$
for all $(R,\vec{a})$  in the Euclidean group. Applying (34) to $\Psi_{0}$, we find
$$
T \Psi_{0} = G_{r}(R,\vec{a}) T \Psi_{0}
\eqno{(35)}
$$
and, since, by lemma 1, $\Psi_{0}$ is the unique vector in ${\cal F}$ invariant under both $G(R,\vec{a})$ and $G_{r}(R,\vec{a})$, (35) yields
$$
T \Psi_{0} = \alpha \Psi_{0}
\eqno{(36)}
$$
where $\alpha$ is a phase. From (32), (33), and (36),
\begin{eqnarray*}
(\Psi_{0}, \Phi(f) \Phi(g) \Psi_{0}) = (\Psi_{0},  T \exp(i\Phi(f))T^{-1} T \exp(i\Phi(g))T^{-1}\Psi_{0})\\
= (\Psi_{0},\exp(i\Phi_{r}(f))\exp(i\Phi_{r}(g))\Psi_{0})
\end{eqnarray*}$$\eqno{(37)}$$
which implies that $\Psi_{0}, \Phi(f) \Phi(g) \Psi_{0})$ and $(\Psi_{0},\exp(i\Phi_{r}(f))\exp(i\Phi_{r}(g))\Psi_{0})$ are equal as tempered distributions on
${\cal S}(\mathbf{R}^{3}) \times {\cal S}(\mathbf{R}^{3})$. We have, from (15), (22),
\begin{eqnarray*}
(\Psi_{0}, \Phi(\vec{x})\Phi(\vec{y})\Psi_{0}) = \frac{1}{2m_{0}} \delta(\vec{x}-\vec{y})=\\
= \frac{1}{(2m_{0})(2\pi)^{3}} \int d\vec{k} \exp(i\vec{k} \cdot (\vec{x}-\vec{y}))
\end{eqnarray*} $$\eqno{(38)}$$
while
\begin{eqnarray*}
(\Psi_{0}, \Phi_{r}(\vec{x}) \Phi_{r}(\vec{y}) \Psi_{0}) = \frac{1}{i}\Delta_{+}(\vec{x}-\vec{y},m_{0}^{2})=\\
=\frac{1}{2(2\pi)^{3}} \int d\vec{k} \exp(i\vec{k}\cdot(\vec{x}-\vec{y}))\frac{c^{2}}{\omega_{\vec{k}}^{c}}
\end{eqnarray*}$$\eqno{(39)}$$
and, by (25),
$$
\frac{c^{2}}{\omega_{\vec{k}}^{c}}= \frac{1}{m_{0}(1+\frac{\vec{k}^{2}}{m_{0}^{2}c^{2}})^{1/2}}
\eqno{(40)}
$$
from which
$$ 
\frac{c^{2}}{\omega_{\vec{k}}^{c}} \to \frac{1}{m_{0}} \mbox{ as } c \to \infty
\eqno{(41)}
$$
For finite $c$, (38) and (39) do not, however, satisfy (37), leading to a contradiction. q.e.d.

In spite of the fact that the relativistic and nonrelativistic zero-time fields lead to inequivalent representations of the CCR due to the fact that the corresponding two-point functions are different for finite $c$, (41) shows that (39) tends to (38) as $c \to \infty$ and suggests that proposal C might be correct This is the content of the forthcoming lemma 2. We assume that we are given two ''wave-functions'' $f_{1},f_{2}$ such that
$$
f_{1}, f_{2} \in {\cal S}(\mathbf{R}^{3})
\eqno{(42)}
$$
\textbf{Lemma 2} Let (42) hold  and $\epsilon $, $\delta $ be chosen such that for $i=1,2$,
\begin{eqnarray*}
\int_{\frac{|\vec{k}|}{m_{0}c}>\delta} d\vec{k}|\tilde{f_{i}}(\vec{k})|^{2} < \epsilon\\
\end{eqnarray*}
$$\eqno{(43)}$$
Then
\begin{eqnarray*}
2m_{0} \Delta C \equiv 2m_{0} |(\Psi_{0}, \Phi(f_{1})\exp(-i(t_{1}-t_{2})H) \Phi(f_{2})\Psi_{0})-\\
- (\Psi_{0}, \Phi_{r}(f_{1}) \exp(-i(t_{1}-t_{2})H_{r}) \Phi_{r}(f_{2})\Psi_{0})| \le\\
\le (2\epsilon + \delta^{2}/2 + |t_{1}-t_{2}|\frac{m_{0}c^{2}}{\hbar} \frac{\delta^{4}}{8})
\end{eqnarray*} $$\eqno{(44)}$$
Above,
$$
H \equiv \int d\vec{k} \frac{\vec{k}^{2}}{2m_{0}} a^{*}(\vec{k})a(\vec{k})
\eqno{(45)}
$$
$$
H_{r} \equiv \int d\vec{k} (\omega_{\vec{k}}^{c}-m_{0}c^{2}) a^{*}(\vec{k})a(\vec{k}) 
\eqno{(46)}
$$
We also define the number operator
$$
N \equiv \int d\vec{k} a^{*}(\vec{k})a(\vec{k})
\eqno{(47)}
$$

\textbf{Remark} It is supposed that $\delta$ is sufficiently small and is coupled to $\epsilon$, so that both are small: a fine tuning is required in (43) and depends on the specific problem, but the requirement (43) is very natural and corresponds to the previously mentioned condition that the wavefunctions are ''small' beyond a certain critical momentum (in the ''relativistic'' region of momenta). In addition the time interval $|t_{1}-t_{2}|$ should be small in comparison with characteristic times related to the rest energy $\frac{\hbar}{m_{0}c^{2}}$ (here we reinserted $\hbar$ for clarity). (45)-(47) may be understood as quadratic forms (see \cite{RSII}, pg. 220). The quantity subtracted in (46) is the ''Zitterbewegungsterm'' \cite{JJS}. Notice that the $2m_{0}$ factor in (44) cancels the product of two $(2m_{0})^{-1/2}$ in each $\Phi(f)$ in (15), or the corresponding relativistic term in the limit $c \to \infty$ by (41).

\textbf{Proof} We write, by (45), (46), (15) and (22), and setting $\tau \equiv t_{1}-t_{2}$,
\begin{eqnarray*}
\Delta C = | \int d\vec{k} (\tilde{f}_{1}(\vec{k})-\tilde{f}_{2}(\vec{k})) \beta(\vec{k},\tau,c)|\\
\beta(\vec{k},\tau,c) \equiv \frac{1}{2m_{0}} \exp(-i\tau \frac{\vec{k}^{2}}{2m_{0}}-\\
- \frac{c^{2}}{\omega_{\vec{k}}^{c}} \exp(-i\tau(\omega_{\vec{k}}^{c}-m_{0}c^{2}))
\end{eqnarray*} $$\eqno{(48)}$$
We split the integral defining $\Delta C$ in (48) into one over $ I_{\delta} \equiv \{\vec{k} ;\frac{|\vec{k}|}{m_{0}c}>\delta \}$, and the other over the complementary region. We now insert the elementary inequalities valid inside $I_{\delta}$: 
\begin{eqnarray*}
\frac{1}{2m_{0}} - \frac{c^{2}}{2\omega_{\vec{k}}^{c}} \le \frac{\delta^{2}}{4m_{0}}\\
|\exp(-i\tau \frac{\vec{k}^{2}}{2m_{0}}) -\exp(-i\tau(\omega_{\vec{k}}^{c}-m_{0}c^{2}))|\\
\le \frac{m_{0}c^{2}|\tau|\delta^{4}}{8\hbar}\\
\frac{c^{2}}{2\omega_{\vec{k}}^{c}} \le \frac{1}{2m_{0}}
\end{eqnarray*}
as well as assumption (43) in the complement of $I_{\delta}$, into (48), to obtain (44). q.e.d.

Lemma 2 shows that in the free field case, in spite of the nonequivalence of the relativistic and nonrelativistic representations shown in theorem 3, Einstein causality is saved, at least in an approximative sense. The real trouble starts with interactions. In that case, (37) implies, taking now for $\Phi$ the interacting field, and $\Psi_{0}$ the interacting vacuum $\Omega_{0}$, that the two-point function of the interacting field must equal that of the free field of mass $m_{0}$ in the case of equivalence of representations. For a hermitian local scalar field for which the vacuum is cyclic, (37) (with $m_{0} > 0$) implies that $\Phi$ is a free field of mass $m_{0}$ (Theorem 4.15 of \cite{StrWight}). We know, however, at least for space dimensions less or equal to 2, interacting fields exist, the first one historically having been in one dimension \cite{GlJa} the free scalar Boson field of mass $m_{0} >0$. Its Hamiltonian is
$$
H(g) = H_{0}+H_{I}(g)= \int_{\mathbf{R}}dk \omega_{k}a^{*}(k)a(k)+\int_{\mathbf{R}}dx g(x):\Phi_{r}(x)^{4}:
\eqno{(49)}
$$
with $g \in L^{2}(\mathbf{R})$ a real valued function. $H(g)$ is a well-defined symmetric operator on a dense set in Fock space (see proposition pg. 227 of \cite{RSII}; for self-adjointness see further in the same reference). The dots in (49) denote the so-called Wick product, which means that all creation operators in $\Phi_{r}(x)^{4}$ are to be placed to the left of all annihilation operators (for further elementary discussion see \cite{MaRo}, and a complete treatment \cite{RSII}). In (49), the limit $g \to \lambda$ (with $\lambda > 0$ a constant, interpreted as the coupling constant) exists in a well-defined sense \cite{GliJa}). In the present case, the vacua $\Omega_{0}$ and the no-particle state, which also belong to inequivalent representations, differ greatly. This may already be expected on the level of (49), because the ground state $\Omega_{g}$ of $H(g)$ (whose existence was proved in \cite{GliJa}) cannot be $\Psi_{0}$ because of the vacuum polarizing term $H_{I}^{P}(g)$ in (49):
$$
H_{I}^{P}(g) = \int dx g(x) \phi_{r}^{*}(x)^{4}
\eqno{(50)}
$$
Those terms in (49) which commute with the number operator $N$ given by (47) are all equal to
$$
H_{I}^{C}(g) = 6 \int dx g(x) \phi_{r}^{*}(x)^{2} \phi_{r}(x)^{2}        
\eqno{(51)}
$$
The formal limit as $c \to \infty$ of the operator $H(g)-H_{I}^{C}(g)$ is not ''small'', for instance for (50) we get from (41)
$$
H_{I,\infty}^{P}(g) = \int dk_{1} \cdots dk_{4} \tilde{g}(k_{1}+\cdots+k_{4})a^{*}(k_{1}) \cdots a^{*}(k_{4})
\eqno{(52)}
$$
in the sense of quadratic forms. In the formal limit $c \to \infty$, $g \to \lambda$, (51) yields
$$
H_{I} = \frac{3\lambda}{2m_{0}^{2}} \int dxdy a^{*}(x)a^{*}(y)\delta(x-y)a(x)a(y)
\eqno{(53)}
$$
with $a^{*}(x),a(x)$ defined by (26): together with $H_{0}$ in (49), this defines the Hamiltonian of a nonrelativistic system of Bosons with delta-function interactions (see \cite{Do} for the precise definition in a segment with periodic b.c.).

The limit $g \to \lambda$, followed by $c \to \infty$, was controlled by Dimock \cite{Dim} in a remarkable tour-de-force. He showed that the two-particle scattering amplitude of model (49) converges to that of model (53) (with the free Hamiltonian (45)). The proof in \cite{Dim} does not, however, offer any hint as to how the contribution of all the terms in $H(g)-H_{I}^{C}(g)$ becomes irrelevant in that limit (W.F.W. thanks Prof. Dimock for a discussion about this topic). 

The above-mentioned point is crucial, for the following reason. For quantum systems in general, it is essential to arrive at many-body systems with \textbf{nonzero} density $\rho$ in the thermodynamic limit, i.e., $N \to \infty$, $V \to \infty$, with $\frac{N}{V}=\rho >0$ (see \cite{MaRo} for an overview of applications). The corresponding non-relativistic system has, in contrast to the situation considered in \cite{Dim}, also an infinite number of degrees of freedom ($N \to \infty$). The situation has an analogy to the classical limit of quantum mechanical correlation functions considered by Hepp \cite{HepC}, where two possible limits may be envisaged, one of them yielding quantum mechanical N-particle systems, the other one classical field theory. For free systems with nonzero density, non-Fock representations arise, both in the non-relativistic and in the relativistic cases \cite{AW}, but it may be checked that lemma 2 continues to hold (for zero temperature). For interacting systems, however, $N$ is not a good quantum number, and, upon fixing it (at a large value proportional to the volume $V$), the relativistic system can only be close to the nonrelativistic one if the contribution of the terms  $H(g)-H_{I}^{C}(g$ becomes indeed irrelevant in the joint limit $g \to \lambda$ followed by $c \to \infty$. 

As an example, we expect that the ground state energy per unit volume of the relativistic system (with a Hamiltonian for volume $V$ defined as in \cite{GliJa}) tends, as $V \to \infty$ and $c \to \infty$, to the thermodynamic limit $e$ of the same quantity in the Lieb-Liniger model \cite{LLi}, which is explicitly known to be $e(\rho)= \rho^{3}f(\frac{\lambda}{\rho})$, with $f$ known explicitly as the unique solution of a Fredholm integral equation. Since $\rho$ is not a parameter in the relativistic system, it is only when the above mentioned terms do not contribute (in the limit $g \to \lambda$, $c \to \infty$) that a similar fixing of the density becomes possible also for the relativistic system. This seems to be a deep mystery, whatever the way the problem is regarded.

In order to explain the last issue more completely, consider the l.h.s. of (44) in lemma 2. In the first term thereof, $\Psi_{0}$ should be replaced by the ground state of the Hamiltonian $H+H_{I}$, where $H$ is given by (45) and $H_{I}$ by (53), and $\Phi(f)$ replaced by a bounded function of the zero time nonrelativistic fields as in (24). Properly speaking, instead of smearing with a function $g$ one should consider the Hamiltonian restricted to a bounded region, e.g. a segment with, say, periodic b.c., and the thermodynamic limit taken, but we shall continue with the previous description for brevity. The second term on the left hand side of (44) should be replaced by
$$
\lim_{g \to \lambda} (\Omega_{g}, A_{r}(f_{1}) \exp(-i(t_{1}-t_{2})H(g)) B_{r}(f_{2})\Omega_{g})
\eqno{(54)}
$$
where $\Omega_{g}$ is the ground state of $H(g)$ (shown to exist in \cite{GliJa}), $A_{r}(f_{1}),B_{r}(f_{2})$ are bounded local functions of the fields (22), (23), i.e., with $f_{1},f_{2}$ with \textbf{compact} support in the space variable. It was shown in \cite{GliJa} that 
$$
\exists\lim_{g \to \lambda} \exp(i \tau H(g)) A(f) \exp(-i \tau H(g))
\eqno{(55)}
$$
in the sense of the norm (in the C*-algebraic sense) for bounded local $A(f)$ (for these concepts, see \cite{BRo2} or \cite{Hug}). This limit, for both operators $A(f_{1}),B(f_{2})$ in (54), determines the observable content of (54), but it is clear that the whole of $H(g)$ will contribute to (55), in particular terms such as
\begin{eqnarray*}
H_{I,c}^{P}(g) = \int dk_{1} \cdots dk_{4} \prod_{i=1}^{4} \frac{c}{(2\omega_{k_{i}}^{c})^{1/2}}\\
\tilde{g}(k_{1}+\cdots+k_{4}) a^{*}(k_{1}) \cdots a^{*}(k_{4})
\end{eqnarray*} $$\eqno{(56)}$$ 
in $H(g)-H_{I}^{C}(g)$ will contribute, for a certain choice of observables in (55). Given that their formal limit as $c \to \infty$ does not vanish ((52)), it seems very unlikely that the limit, as $c \to \infty$, of (55) is independent of $H(g)-H_{I}^{C}$. In this connection, one may recall that the $S$-matrix, considered in \cite{Dim} as an observable, is of a special kind, because it commutes with the free Hamiltonian $H_{0}$ in (49).

The basic ingredient of the proof of (55) \cite{GliJa} is the fundamental property of microcausality (1) \cite{Haag}\cite{StrWight}. On the other hand, the form (49) of the Hamiltonian is dictated by the property of Lorentz covariance \cite{GlJa}, proved for this model in \cite{HeOs}.

\section{Conclusion}

In this review we discussed two aspects of the dynamics of non-relativistic quantum systems, unified by a dichotomy in Hegerfeldt's theorem 1. According to this theorem, there are exactly two options (5) and (6) for such systems.

The first aspect was related to option (6) in that theorem, viz. the attempt to isolate a quantum system from its surroundings by a set of boundary conditions, including those of Dirichlet and Neumann type (a1,a2). In general, this leads to physical inconsistencies, as reviewed in theorem 2.2 for non-relativistic systems, see also \cite{GarKar}. We view these inconsistencies as consequences of trying to impose conditions deriving from classical physics to quantum systems, be they non-relativistic or relativistic. The latter case is well illustrated in Milonni's famous paper \cite{Mil}, where he proved that, near a perfectly reflecting slab, the transverse vector potential and the electric field satisfy a set of equal-time CCR different from those holding for free fields. In \cite{BohrRos}, Bohr and Rosenfeld showed, under the natural assumption that the fields are measured by observing the motion of quantum massive objects with which the fields interact, that the above-mentioned equal-time CCR follow. They are, therefore, very fundamental.  

This suggests that such idealized b.c. are unphysical: this fact was explicitly shown for Dirichlet or Neumann b.c. in the case of the Casimir effect \cite{KNW}. The reason is that there are wild fluctuations of quantum fields over sharp surfaces \cite{DeCa}. One promising direction to study this (as yet open) problem is to look at the electromagnetic field in the presence of dielectrics, instead of the ''infinitely thin'' conductor plates (see \cite{Bim} and references given there). The Casimir problem (for conductors or dielectrics) is an example for which adoption of generalized periodic b.c. (a3) in theorems 2.1,2.2 is not a physically reasonable option: it is not compatible with the theory's classical limit.

Our second topic concerned option (5) in theorem 1, the issue of instantaneous spreading for non-relativistic quantum systems. We concluded that there are serious obstacles on the way to rescue Einstein causality in the (natural) approximative sense of lemma 2, due to terms such as the vacuum polarizing term (56) in the interaction Hamiltonian (49), which are not ''small'' in the formal limit $c \to \infty$ by (52). Since the presence of such terms is dictated by such fundamental principles as Lorentz covariance and microcausality, the solution may not be simple.

Although we used a special model for the sake of argument, any physical theory with vacuum polarization, such as quantum electrodynamics, is expected to be subject to analogous considerations. Notice that the remarks on the use of approximate theories in \cite{Yng} do not apply here, because the problems we pose are not due to the approximative character of the theories, such as, e.g., various cutoffs in quantum-electrodynamics, but, as remarked in the previous paragraph, are due to an intrinsic property of relativistic quantum field theory, viz., vacuum polarization (or, more precisely, having a non-persistent vacuum).

The arguments we presented, however, are clearly no mathematical proof of a no-go theorem. One reason is that the limits $g \to \lambda$ and $c \to \infty$ do not necessarily commute: they may not. The problem is therefore open. It is hoped that a complete change of point of view may clarify the problem, but we conjecture that  $H(g)-H_{C}^{I}(g)$ will play a central role in the final solution.   

Progress in both topics above would obviously be of great relevance for the foundations of quantum theory.   

\textbf{Acknowledgement} The idea of a part of this review arose at the meeting of operator algebras and quantum physics, satellite conference to the XVIII international congress of mathematical physics. We thank the organizers for making the participation of one of us (W. F. W.) possible, and Prof. J. Dimock for discussions there on matters related to section 3. We also thank Christian J\"{a}kel for critical remarks concerning possible changes of viewpoint, and for recalling some relevant references. W.F.W. also thanks J. Froehlich for calling his attention to the reference \cite{Yng}.

\end{document}